\documentclass{ws-ijmpcs}

\begin{document}

\markboth{Peralta-Ramos, J., Nakwacki, M.S.}{Production of thermal photons in a simple chiral-hydrodynamic model}

%
\catchline{}{}{}{}{}
%

\title{PRODUCTION OF THERMAL PHOTONS IN A SIMPLE CHIRAL-HYDRODYNAMIC MODEL}

\author{PERALTA-RAMOS, J.}

\address{Instituto de F\'isica Te\'orica, Universidade Estadual Paulista \\
Rua Doutor Bento Teobaldo Ferraz, 271 - Bloco II, S\~ao Paulo, SP, Brazil\\
jperalta@ift.unesp.br}

\author{NAKWACKI, M.S.}

\address{Instituto de Astronom\'ia, Geof\'isica e Ci\^encias Atmosf\'ericas, Universidade de S\~ao Paulo, \\ 
Rua do Mat\~ao 1226, Cidade Universit\'aria, 05508-090 S\~ao Paulo, Brazil\\
sole@astro.iag.usp.br}

\maketitle

\begin{history}
\received{Day Month Year}
\revised{Day Month Year}
\end{history}

\begin{abstract}

We use a self-consistent chiral-hydrodynamic formalism which combines the linear $\sigma$ model with second-order hydrodynamics in $2+1$ dimensions to compute the spectrum of thermal photons produced in Au+Au collisions at $\sqrt{s_{NN}}=200$ GeV. The temperature-dependent shear viscosity of the model, $\eta$, is calculated from the linearized Boltzmann equation. We compare the results obtained in the chiral-hydrodynamic model to those obtained in the second-order theory with a Lattice QCD equation of state and a temperature-independent value of $\eta/s$. We find that the thermal photon production is significantly larger in the latter model due to a slower evolution and larger dissipative effects. 

\end{abstract}

\ccode{PACS numbers: 25.75.-q, 24.10.Nz, 24.85.+p}

\section{Introduction}


In this paper we focus on the spectrum of thermal photons produced during the expansion of the fireball created in Au+Au collisions at $\sqrt{s_{NN}}=200$ GeV. Photons created in the interior of the fireball pass through it without any interaction, thus giving information on the properties of {\it bulk} nuclear matter \cite{rev,phenix,nos,nosMPLA,bhat,dus}.
Thermal photon spectra have been calculated within the framework of different fluid dynamics models \cite{rev,dus,bhat,nos,nosMPLA} as well as a hybrid kinetic-hydrodynamic model \cite{bleich}. 
Here we employ a self-consistent chiral-hydrodynamic model combining second-order (SOT) hydrodynamics and the linear $\sigma$ model (LSM), in which the temperature-dependent $\eta/s$ is computed from kinetic theory \cite{jeroprc}. We compare the spectrum of thermal photons with that obtained from the same SOT with a Lattice QCD-inspired equation of state (EOS) and a temperature-independent value of $\eta/s$. The latter model is commonly employed in the study of hadronic observables of heavy ion collisions. The purpose of this contribution is to quantify the differences in thermal photon spectra that arise from differences in these two fluid dynamical models when they are taken as a whole. In other words, no attempt is made to disentangle the impact of EOS or the temperature dependence of $\eta/s$ on the thermal photon spectra (see \cite{nosMPLA} for a study along this line). 
 


\section{Chiral-hydrodynamic model}

We now describe the coupling between the chiral fields and the quark fluid, the latter described by second order boost invariant fluid dynamics (see \cite{nos,nosMPLA,jeroprc} for further details). 


The classical equations of motion of the LSM are given by 
\begin{equation}
  D_\mu D^\mu \phi_a + \frac{\delta U}{\delta \phi^a} = -g\rho_a 
\label{eomchiral}
\end{equation}
where $\phi=(\sigma,\vec{\pi})$ and 
\begin{equation}
  \rho_a = g \phi_a d_q \int \frac{d^3 p}{(2\pi)^3} \frac{1}{\sqrt{p^2 + g^2 \sum_a \phi_a^2 }} f_0(p^\mu,x^\mu) 
\label{rhosrhops}
\end{equation}
$f_0(p^\mu,x^\mu)$ is the Fermi--Dirac function and $U$ is the potential exhibiting chiral symmetry breaking. These are solved selfconsistently together with the conservation equations for the combined system of chiral fields plus quark fluid  
\begin{equation}
\begin{split}
D\epsilon &= -(\epsilon+p)\nabla_\mu u^\mu + \Pi^{\mu\nu}\sigma_{\mu\nu} + g (\rho_s D \sigma + \vec{\rho}_{ps}\cdot D\vec{\pi}) \\
(\epsilon+p)Du^i &= \frac{1}{3}(g^{ij}\partial_j \epsilon - u^i u^\alpha \partial_\alpha \epsilon) - \Delta^i_\alpha D_\beta \Pi^{\alpha\beta} + g (\rho_s \nabla^i \sigma + \vec{\rho}_{ps}\cdot \nabla^i\vec{\pi}) ~. 
\end{split}
\label{conseq}
\end{equation}
where $(\rho_s,\vec{\rho}_{ps})=\rho_a$, $D=u_\mu D^\mu$ is the comoving time derivative,   
$\nabla_\mu = \Delta_{\mu\alpha}D^\alpha$ is the spatial gradient, and $\sigma_{\mu\nu}$ is shear tensor. 
The evolution equation for $\Pi^{\mu\nu}$ contains terms up to second order in velocity gradients and it is given in \cite{nos,nosMPLA}.
In order to estimate the temperature dependence of the shear viscosity $\eta(T)$ in the LSM, we adapt the method described in \cite{sas} and employ the linearized Boltzmann equation in the relaxation time approximation (see \cite{jeroprc}).


To compute the spectrum of thermal photons created during the evolution of the fireball we consider the processes of Compton scattering, $q\bar{q}$ annihilation and bremsstrahlung in the QGP phase, and $\pi\pi \rightarrow \rho \gamma$, $\pi\rho \rightarrow \pi \gamma$ and $\rho \rightarrow \pi \pi \gamma$ in the hadron phase \cite{rev}, as described in \cite{nos}. In all cases, the 
nonequilibrium distribution function of the quarks, $f(x^\mu,p^\mu)$, is calculated from Grad's ansatz. 
\section{Results}
\label{res}
We now go over to present our results. As a typical example of a conformal fluid we take $\tau_\pi = 2(2-\ln 2)\frac{\eta}{sT}$ and $\lambda_1=\frac{\eta}{2\pi T}$ corresponding  to the supersymmetric Super--Yang--Mills plasma. We use a 13 fm $\times$ 13 fm transverse plane, and set the impact parameter to  $b=7$ fm. The initialization time is set to $\tau_0 = 1$ fm/c, and we use  $u^x=u^y=0$ and $(\Pi^{xx},\Pi^{xy},\Pi^{yy})=0$ as initial values. The initial energy density profile is obtained from Glauber's model, with a temperature $T_i=330$ MeV at the center of the fireball. We take $\vec{\pi}(\tau_0,\vec{r})=0$ and $\sigma(\tau_0,\vec{r}) = f_\pi [1-e^{-(r/r_0)^2}]$ with $r_0 = 9$ fm as initial values for the chiral fields. We use the isothermal Cooper-Frye freeze--out prescription with a freeze-out temperature $T_F = 130$ MeV. The critical temperature is set to $T_c = 170$ MeV.

The upper panel of Figure \ref{cs} shows the square of the speed of sound $c_s^2=\partial p/\partial \epsilon$ corresponding to the LSM with $g=3.2$, which yields a smooth crossover, and to the EOS calculated by Laine and Schr\"{o}der \cite{laine} which connects a high-order weak-coupling perturbative QCD calculation at high temperatures to a hadron resonance gas at low temperatures via an analytic crossover.  Recent works have shown that the comparison of hydrodynamic simulations with RHIC data favor a smooth crossover over a first-order transition \cite{cspaper}, and this is the reason for considering a crossover (and not a first-order phase transition) in the LSM. There is no appreciable change in the photon spectrum provided $g\sim 3.2$, but if $g>3.4$ or $g<3$ the dip in $c_s^2$ at $T_c$ becomes too sharp or fades away, respectively, resulting in a temperature dependence that strongly disagrees with the one obtained in Lattice QCD calculations. As it can be seen from Figure \ref{cs}, for large temperatures the conformal limit $c_s^2=1/3$ is reached, while for low values $c_s^2$ goes to zero. Note also that the transition region is significantly broader in the LQCD EOS than in the LSM.
\newcommand{\imsize}{1.0\columnwidth}
\begin{figure}
\begin{center}
{\resizebox{\imsize}{!}{\includegraphics{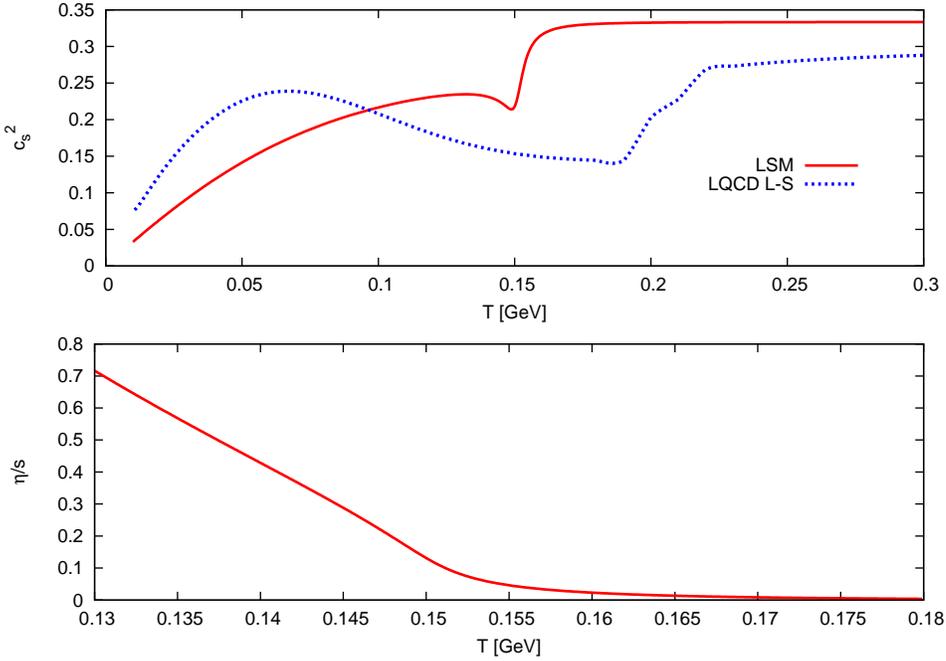}}}
\vspace{-0.2cm}
\caption{(Color online) Speed of sound of the LSM and that of the model EOS of Laine and Schr\"{o}der$^{15}$  ({\it upper panel}); $\eta(T)/s(T)$ in the LSM ({\it lower panel}). The value of the coupling constant is $g=3.2$.}\label{cs}
\end{center}
\vspace{-0.3cm}
\end{figure}

The lower panel of Figure \ref{cs} shows $\eta/s$ as a function of temperature for the LSM with $g=3.2$. It is seen that the ratio remains small ($< 0.1$) in the QGP phase and starts increasing abruptely for $T\lesssim T_c$. For this reason we must impose a cut-off in the value of $\eta/s$; we take \footnote{Any hydrodynamic theory necessarily breaks down when viscosity is large enough. The value of $\eta/s$ at which it is sensible to impose this cut-off can be estimated in $(\eta/s)|_c \sim 0.4-0.5$ (see e.g. \cite{jeroprc}).} $\eta/s \leq 0.4$. For the SOT with the LQCD EOS we consider a temperature-independent value $\eta/s = 0.11$, which corresponds to the averaged value throughout the evolution.

\begin{figure}
\begin{center}
{\resizebox{\imsize}{!}{\includegraphics{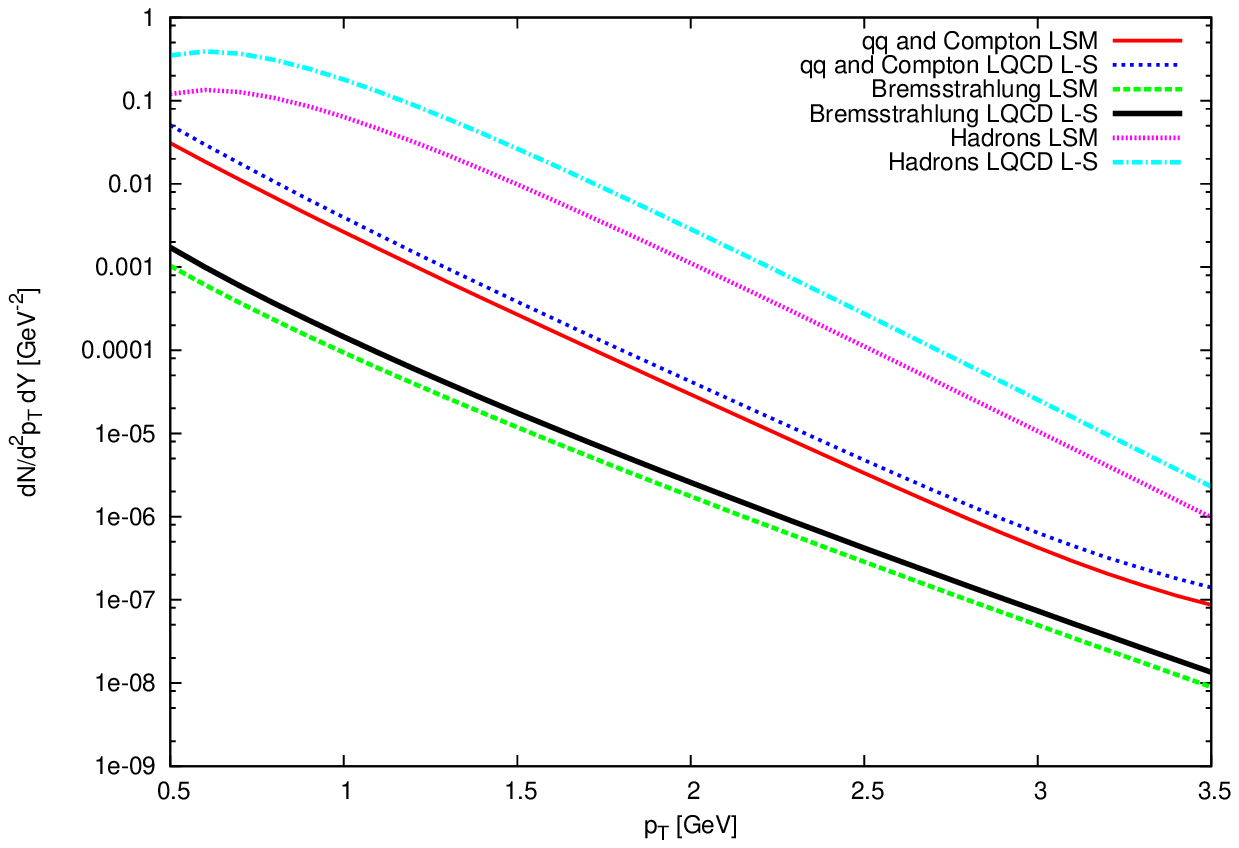}}}
\vspace{-0.2cm}
\caption{(Color online) Contributions to the thermal photon spectra from the QGP and hadronic phases for the LSM with $g=3.2$ and the SOT with Lattice QCD EOS.}\label{qgp}
\end{center}
\vspace{-0.3cm}
\end{figure}

Figure \ref{qgp} shows the different contributions to the total thermal photon spectrum coming from the QGP and the hadronic phases, for the LSM and the SOT with the Lattice QCD EOS. 
It is seen that in all cases the photon spectrum is considerably smaller in the LSM, the difference being larger in the hadronic phase. The difference between the spectra in both models is practically independent of $p_T$, except at very low values where the difference becomes slightly larger. In both models, the contribution of the hadronic phase to the photon spectrum is dominant over that coming from the QGP phase.

There are two reasons that make the spectrum obtained from the chiral model smaller than the corresponding to the Lattice QCD EOS. First, the hydrodynamic evolution is much faster in the LSM model, essentially because the speed of sound is on average larger than the one corresponding to the Lattice QCD EOS. This is illustrated by the fact that the  freeze-out temperature is reached in $\sim 6$ fm/c in the LSM and in $\sim 10$ fm/c, which represents a significant difference. The other reason is that the shear tensor is on average significantly smaller in the LSM, leading to a smaller contribution of dissipative effects to the total photon spectrum.

\section{Conclusions}
\label{conc}
Using a simple self-consistent chiral-hydrodynamic model based on the linear $\sigma$ model and $2+1$ conformal second-order hydrodynamics, we calculated the spectrum of thermal photons including nonequilibrium corrections produced during the evolution of the fireball created in Au+Au ultrarelativistic collisions. Compared to the results obtained with a commonly used second-order fluid dynamic model with an equation of state inspired in Lattice QCD and a temperature independent $\eta/s$, we find that the photon production is significantly smaller in the chiral-hydrodynamic model, due to a faster evolution and to  smaller dissipative corrections to the nonequilibrium distribution function.

\section*{Acknowledgements}

The authors acknowledge FAPESP (S\~ao Paulo, Brazil) for financial support.

\end{document}